\documentclass[letterpaper, 10 pt, conference]{ieeeconf} 

\IEEEoverridecommandlockouts 

\usepackage{placeins}
\usepackage{float}
\usepackage{graphics} 
\usepackage{epsfig} 
\usepackage{mathptmx} 
\usepackage{times} 
\usepackage{amsmath} 
\usepackage{amssymb} 
\usepackage{algorithm} 
\usepackage{verbatim} 
\usepackage{algpseudocode}
\usepackage{array}
\usepackage{epsfig}
\usepackage{amsmath}
\usepackage{float}
\usepackage{graphicx}
\usepackage{tabto}
\usepackage{cleveref}
\usepackage{subfigure}
\usepackage{wrapfig}
\usepackage{textcomp}
\usepackage{float}
\usepackage{booktabs}
\usepackage{graphicx}
\usepackage{url} 
\usepackage{multirow}
\usepackage{amsmath}
\usepackage{flexisym}
\usepackage{cite}
\usepackage{filecontents}
\usepackage{caption}

\title{\LARGE \bf
Automatic Frame Selection using CNN in Ultrasound Elastography
\\
}

\author{Abdelrahman Zayed$^{1}$, Student member, IEEE, Guy Cloutier $^{2}$ and Hassan Rivaz$^{3}$, Senior members, IEEE
\thanks{*This research was funded by Richard and Edith Strauss Foundation and the Quebec Bio-imaging Network (QBIN).}
\thanks{$^{1}$Abdelrahman Zayed is with Department of Electrical and Computer Engineering
and PERFORM Centre, Concordia University, Montreal, Quebec, Canada
{\tt\small a\_zayed@encs.concordia.ca}}%
\thanks{$^{2}$Guy Cloutier is with Department of Radiology, Radio-oncology and Nuclear Medicine, University of Montreal, Montreal, Quebec, Canada
{\tt\small guy.cloutier@umontreal.ca}}%
\thanks{$^{3}$Hassan Rivaz is with Department of Electrical and Computer Engineering
and PERFORM Centre, Concordia University, Montreal, Quebec, Canada
{\tt\small hrivaz@ece.concordia.ca}}%
}

\begin{document}

\maketitle
\thispagestyle{empty}
\pagestyle{empty}

\begin{abstract}
Ultrasound elastography is used to estimate the mechanical properties of the tissue by monitoring its response to an internal or external force. Different levels of deformation are obtained from different tissue types depending on their mechanical properties, where stiffer tissues deform less. Given two radio frequency (RF) frames collected before and after some deformation, we estimate displacement and strain images by comparing the RF frames. The quality of the strain image is dependent on the type of motion that occurs during deformation. In-plane axial motion results in high-quality strain images, whereas out-of-plane motion results in low-quality strain images. In this paper, we introduce a new method using a convolutional neural network (CNN) to determine the suitability of a pair of RF frames for elastography in only 5.4 $ms$. Our method could also be used to automatically choose the best pair of RF frames, yielding a high-quality strain image. The CNN was trained on 3,818 pairs of RF frames, while testing was done on 986 new unseen pairs, achieving an accuracy of more than $91 \%$. The RF frames were collected from both phantom and \textit{in vivo} data. 

\end{abstract}

\begin{figure*}[h]
\begin{center}
\centering

\subfigure[Data collection]{\includegraphics[height=5.3 cm,width=5.5 cm]{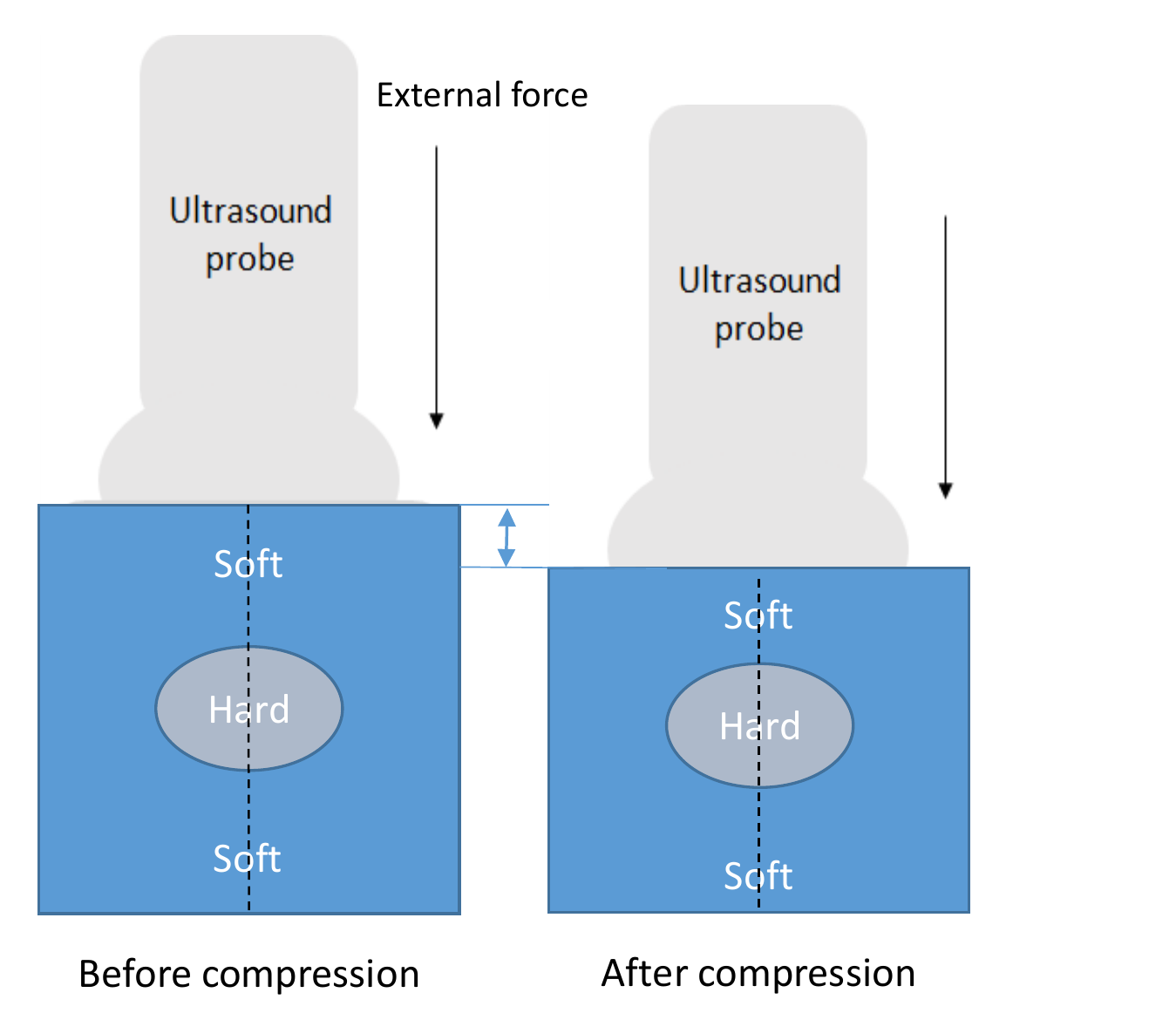}}\hspace*{-0em}
\subfigure[Displacement]{\includegraphics[height=3.5 cm,width=3.5 cm]{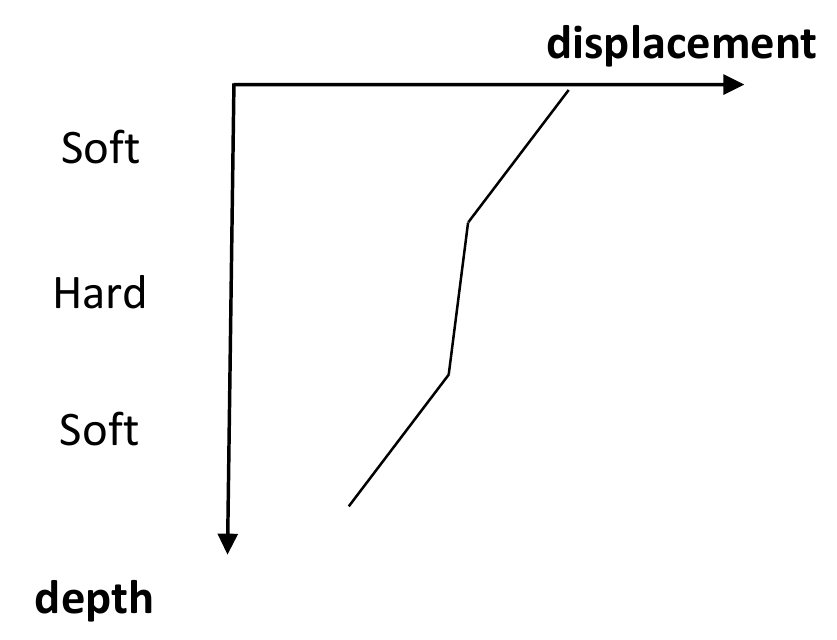}}\hspace*{-0em}
\subfigure[Strain]{\includegraphics[height=3.5 cm,width=3.5 cm]{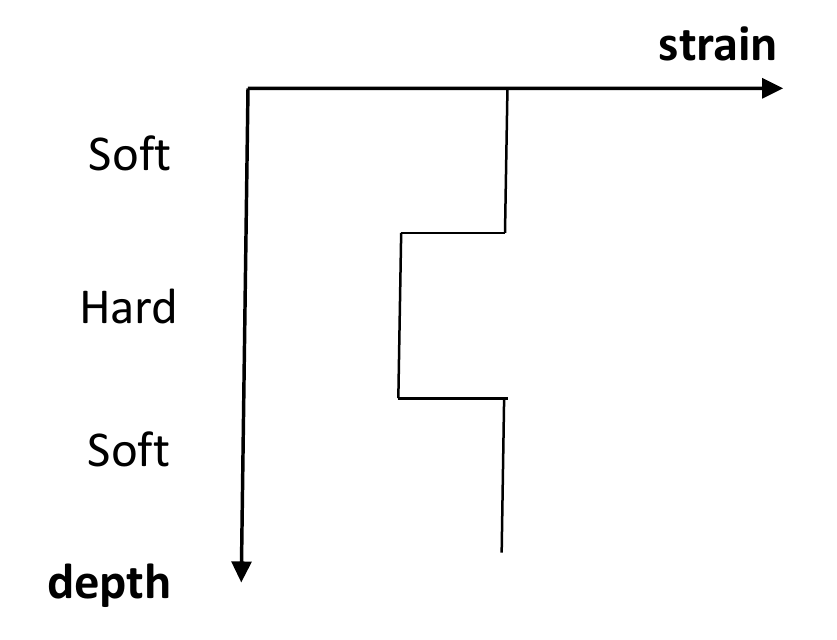}}\hspace*{-0em}

\end{center}%
\caption{The basic steps of quasi-static ultrasound elastography. After data collection in (a), we estimate the amount of displacement in every sample in the RF frame, yielding the displacement image as shown in (b). Finally, we obtain the strain image in (c) by spatially differentiating the displacement image.} \label{elastography}
\end{figure*}

\begin{figure*}[h]
\captionsetup{justification=centering}
\begin{center}
\includegraphics[height=7 cm,width=17 cm]{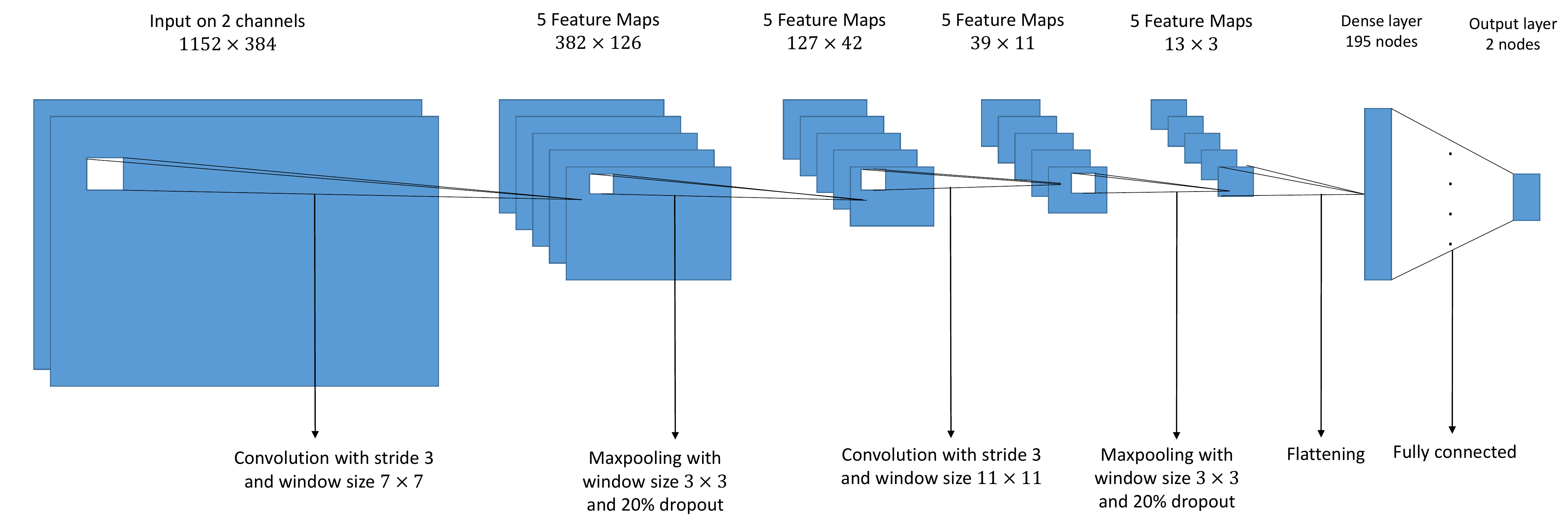}
\end{center}%
\centering
\caption{The architecture of the CNN used for RF frame selection.}%
\label{CNN_architecture}%
\end{figure*}

\begin{figure*}[h!]
\begin{center}
\centering
\subfigure[B-mode]{\includegraphics[height=5.3 cm,width=4.2 cm]{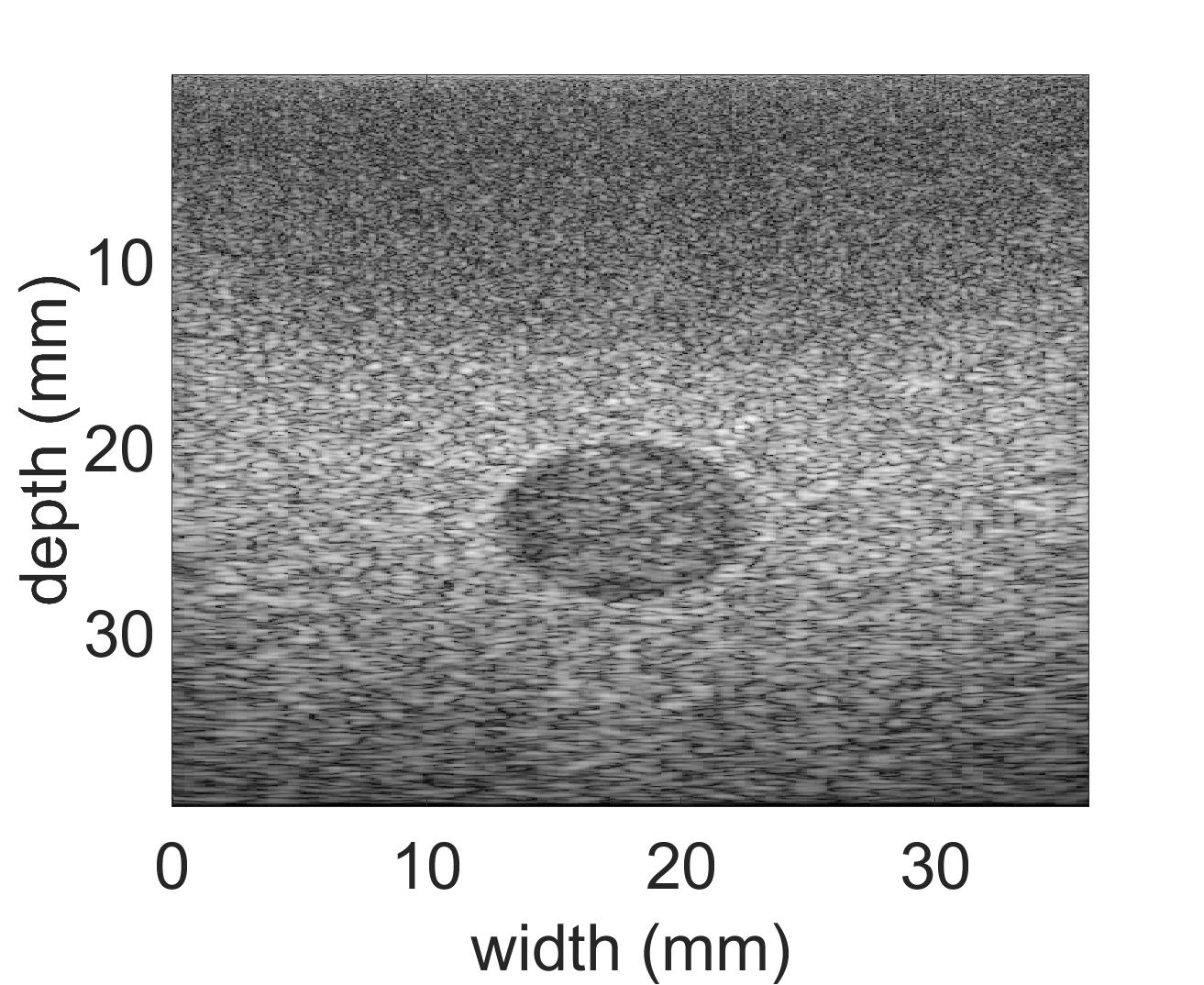}}\hspace*{-0em}
\subfigure[Strain from our method]{\includegraphics[height=5.3 cm,width=4.2 cm]{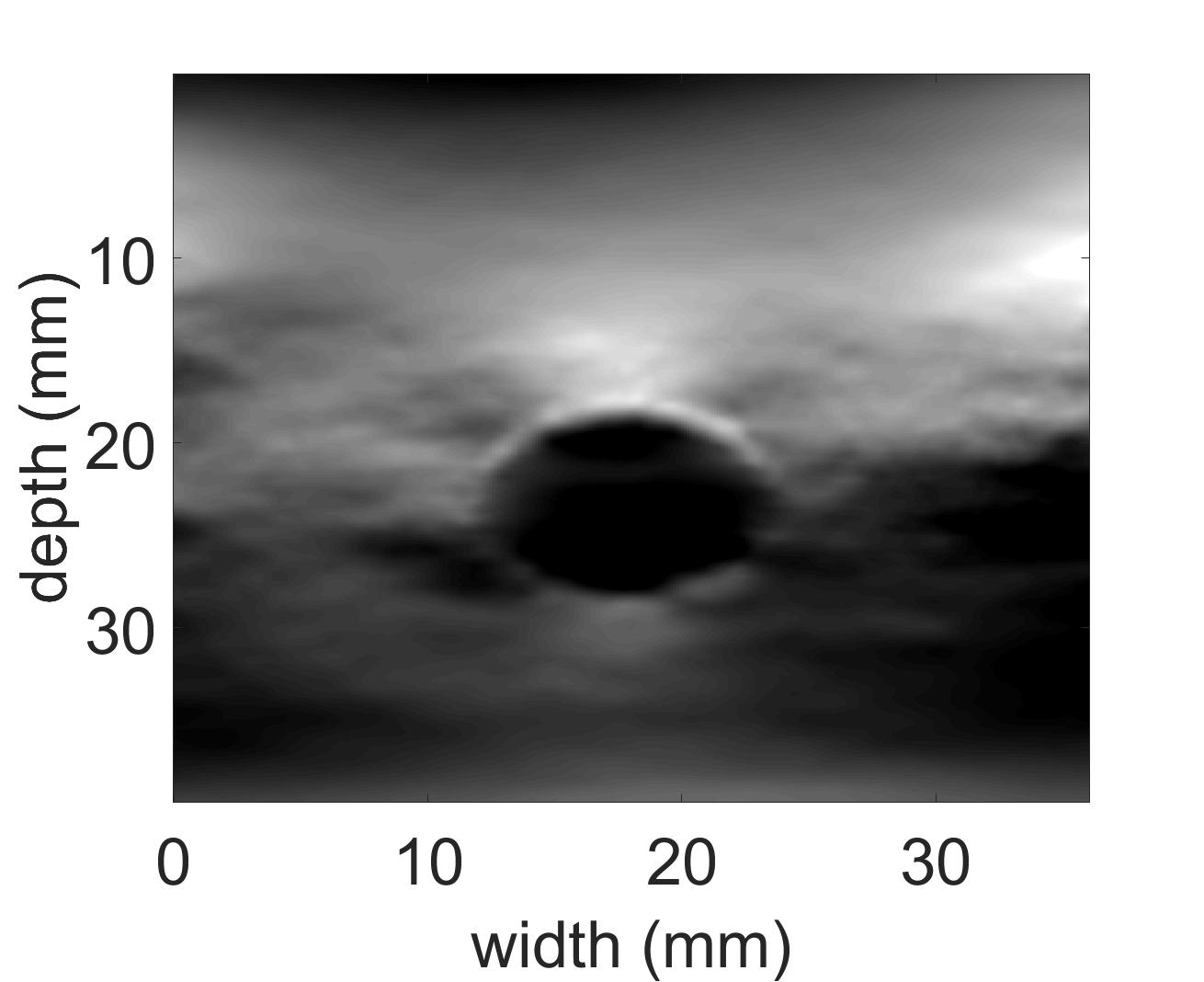}}\hspace*{-0em}
\subfigure[Strain from Skip 1 method]{\includegraphics[height=5.3 cm,width=4.2 cm]{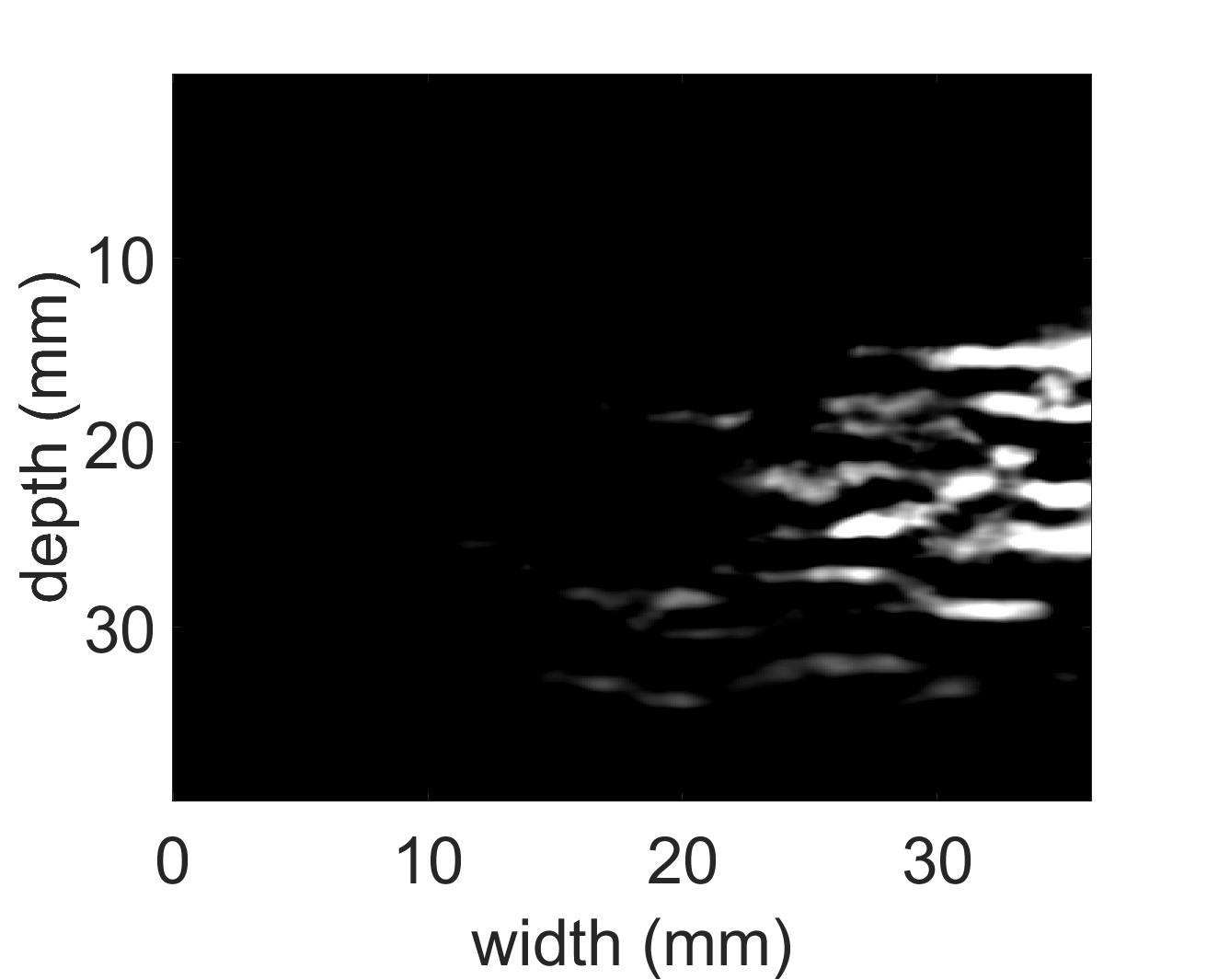}}\hspace*{-0em}
\subfigure[Strain from Skip 2 method]{\includegraphics[height=5.3 cm,width=4.2 cm]{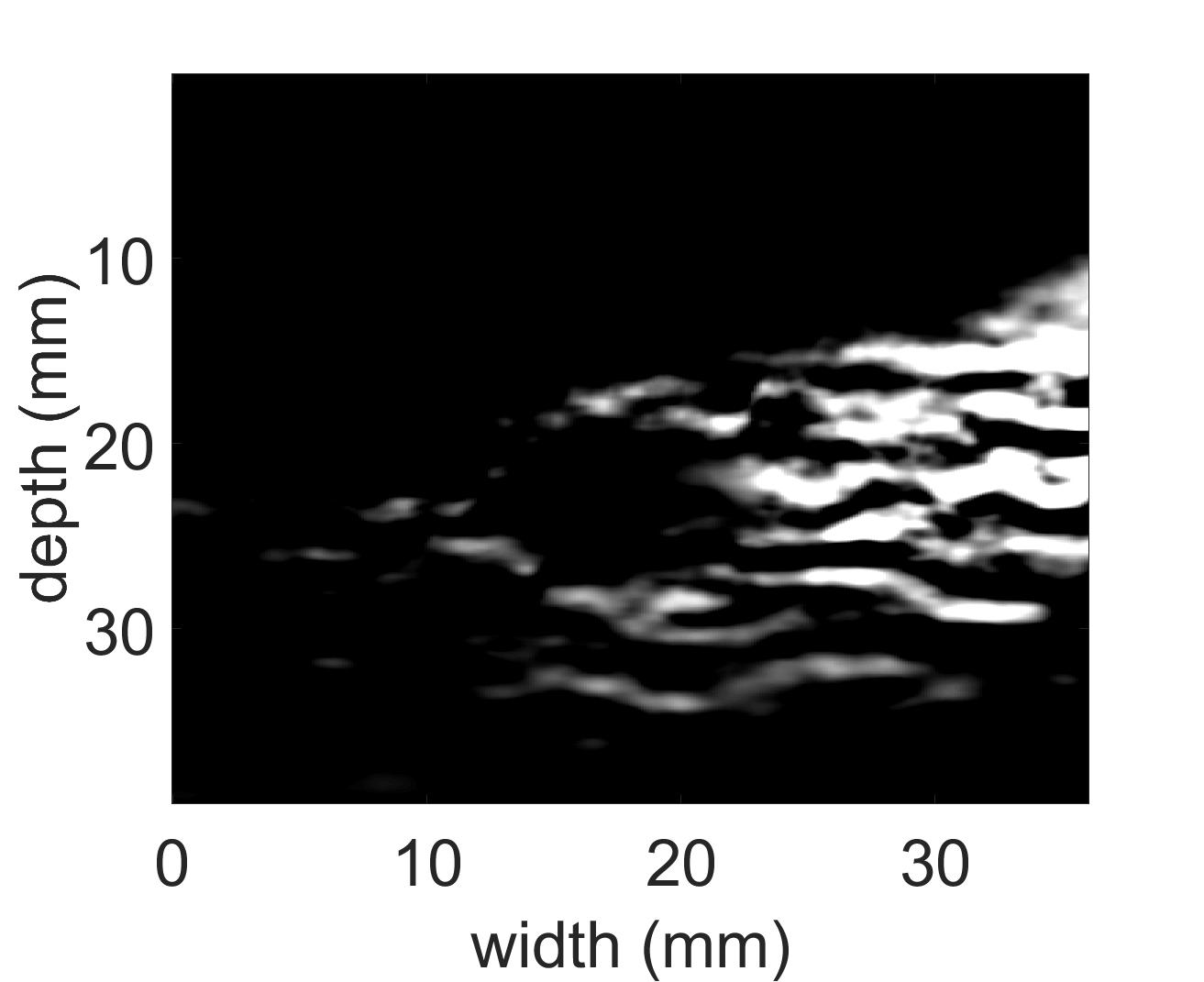}}\vspace*{-0em}
\subfigure[Color bar]{\includegraphics[height=1.2 cm,width=4.2 cm]{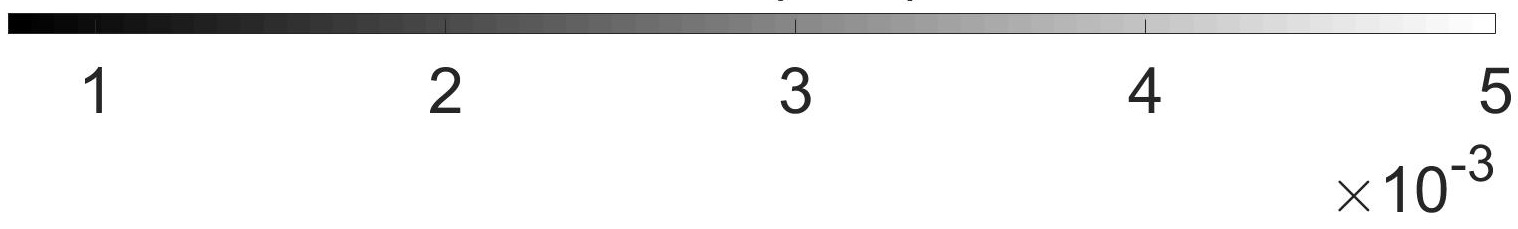}}\vspace*{-0em}
\end{center}
\caption{The B-mode ultrasound and PCA-GLUE axial strain image for the phantom experiment using different frame selection methods. Poor strain images in (c) and (d) are rejected by the proposed method. The color bar is for the strain images.} \label{phantom_results}
\end{figure*} 

\begin{figure*}[h!]
\begin{center}
\centering

\subfigure[B-mode]{\includegraphics[height=5.3 cm,width=4.2 cm]{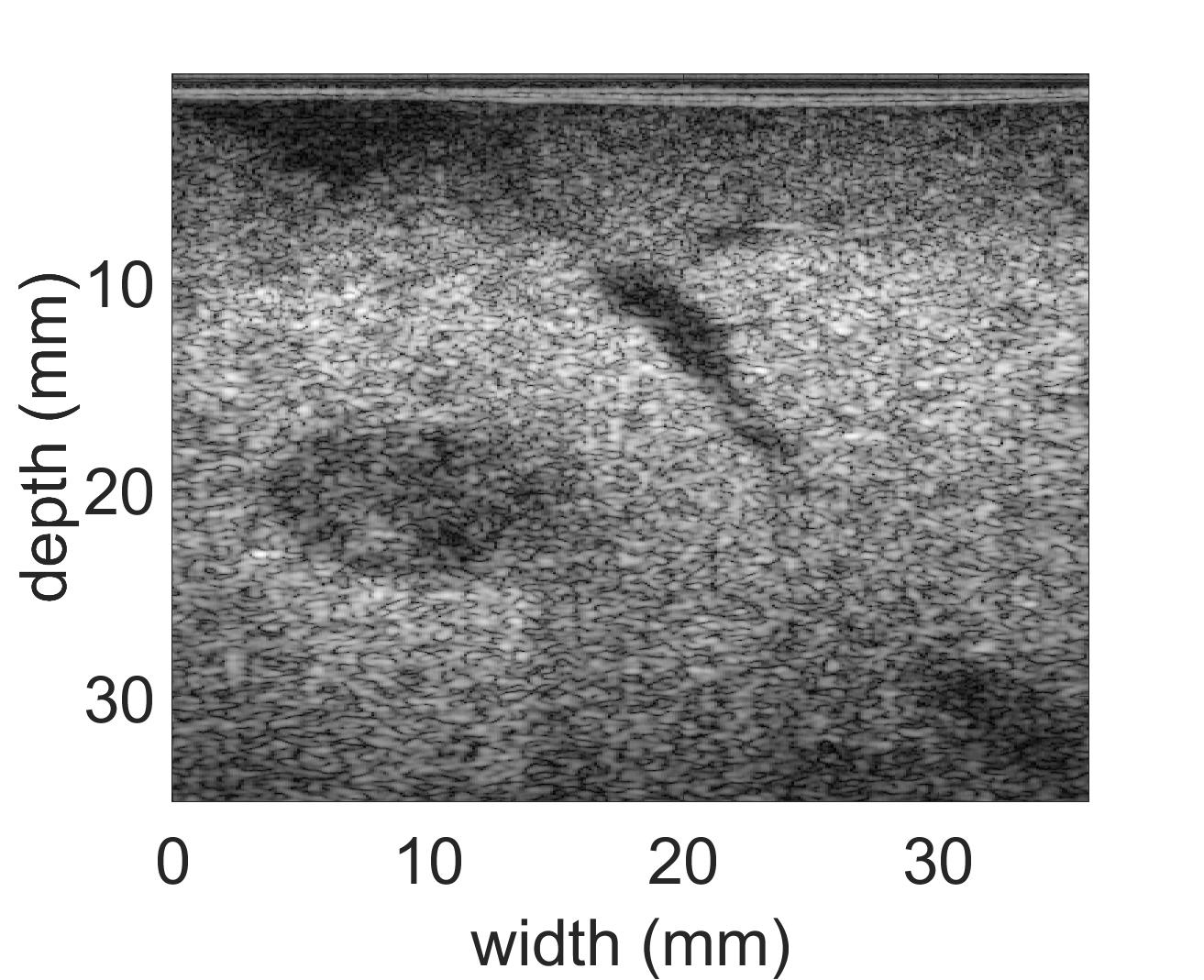}}\hspace*{-0em}
\subfigure[Strain from our method]{\includegraphics[height=5.3 cm,width=4.2 cm]{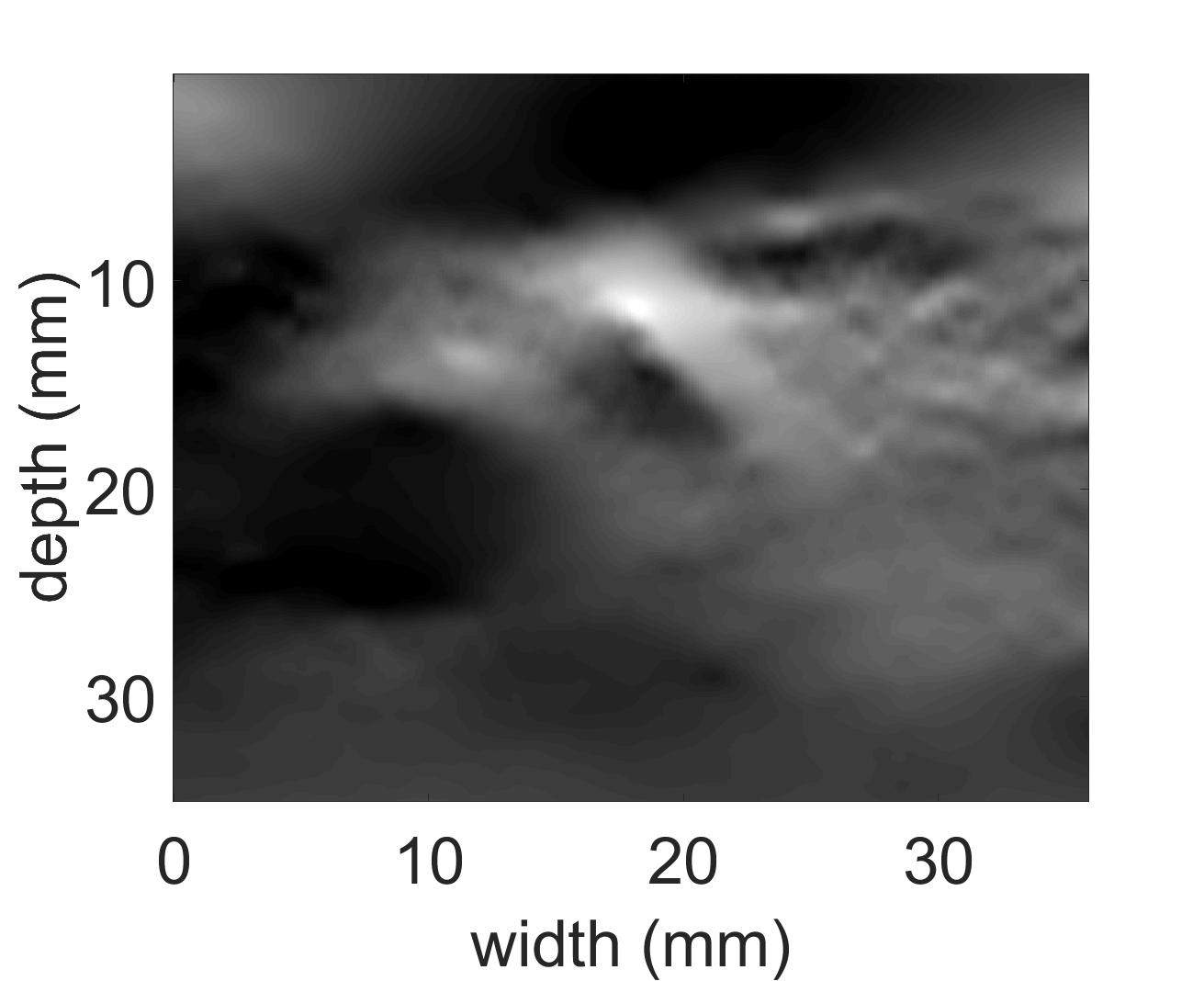}}\hspace*{-0em}
\subfigure[Strain from Skip 1 method]{\includegraphics[height=5.3 cm,width=4.2 cm]{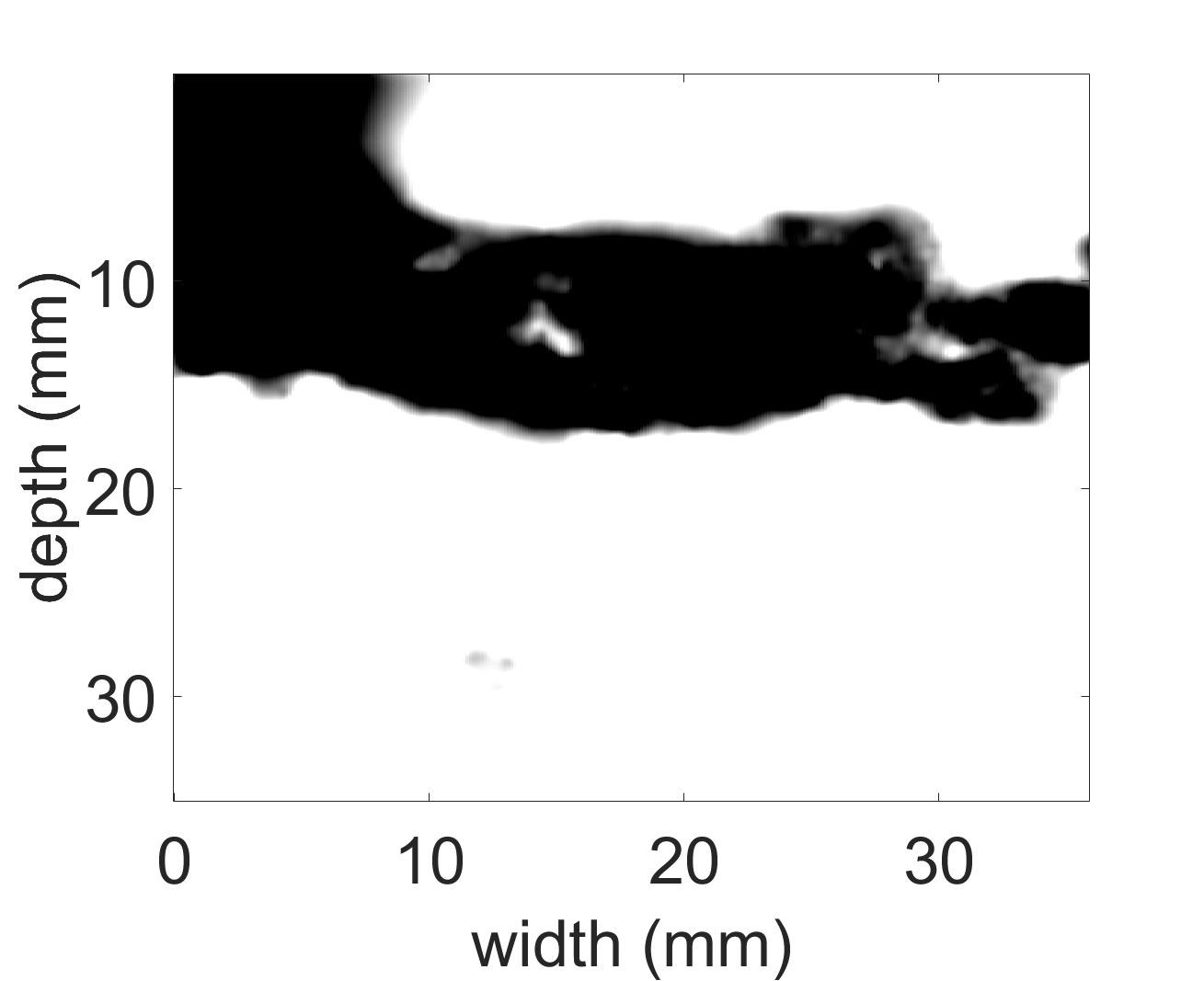}}\hspace*{-0em}
\subfigure[Strain from Skip 2 method]{\includegraphics[height=5.3 cm,width=4.2 cm]{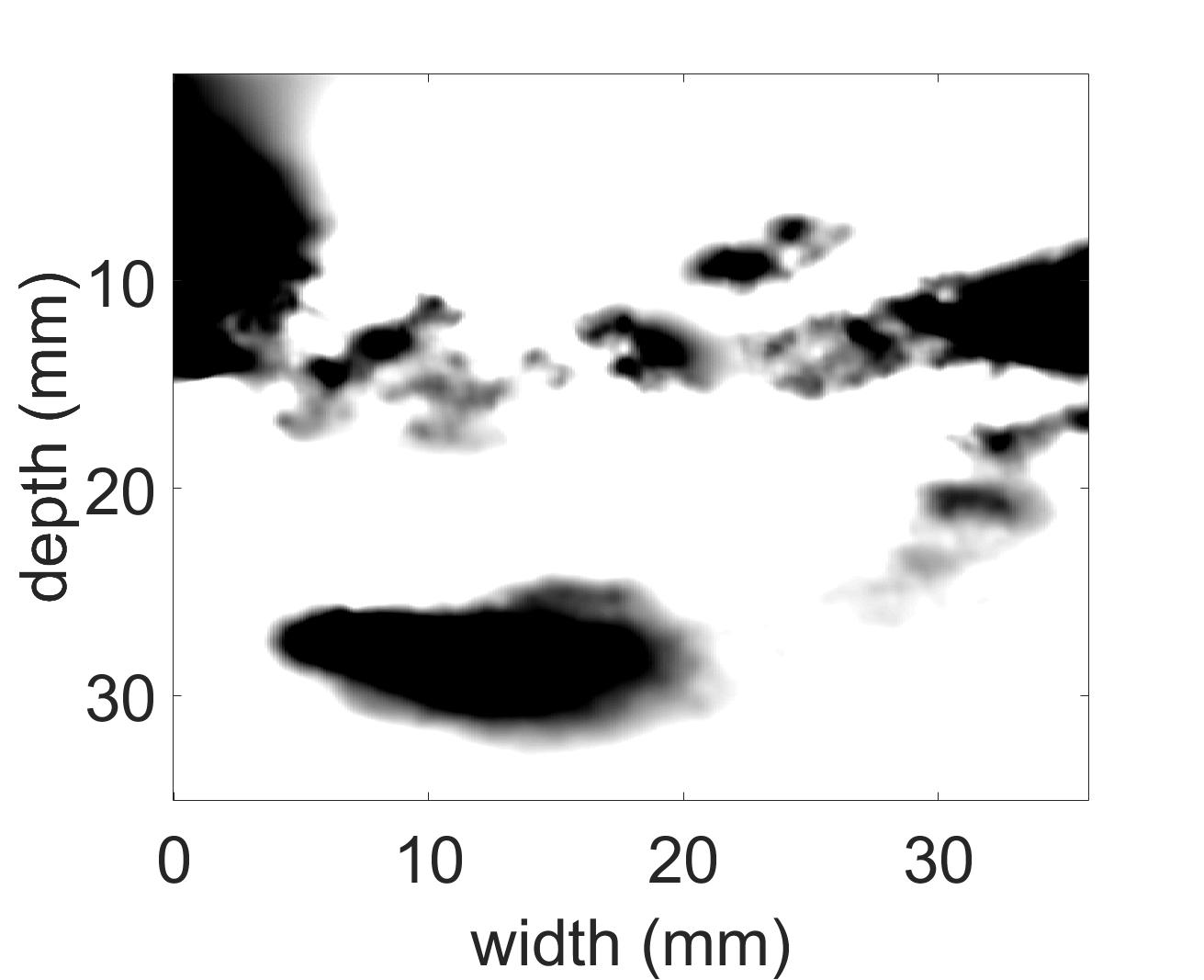}}\vspace*{-0em}
\subfigure[Color bar]{\includegraphics[height=1.2 cm,width=4.2 cm]{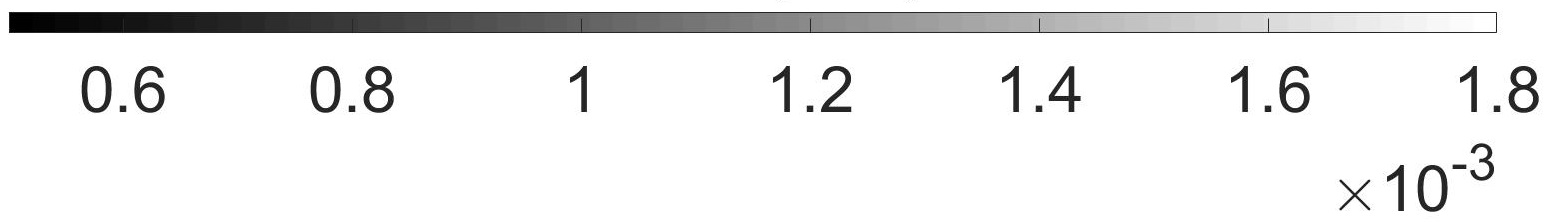}}\vspace*{-0em}

\end{center}%
\caption{The B-mode ultrasound and PCA-GLUE axial strain image for the \textit{in vivo} experiment using different frame selection methods. The low-quality strain images in (c) and (d) are rejected by the proposed method.  The color bar is for the strain images.} \label{invivo_results}
\end{figure*} 

\section{INTRODUCTION}


\label{sec:intro}
Ultrasound has numerous applications in the diagnosis and treatment of different diseases. Ultrasound elastography is a branch of ultrasound that studies the mechanical properties in the tissue such as strain. A detailed review of elastography and its clinical applications can be found at~\cite{brian_anthony,app1,app2,application}.


Ultrasound elastography can be classified into two types of quasi-static or dynamic \cite{j2011recent}. In the first type, the deformations are very slow and therefore tissue dynamics can be ignored \cite{parker2010imaging,ophir1999elastography,treece2011real}. Freehand quasi-static imaging does not need any
additional hardware and as such, is very common (Fig.~\ref{elastography}). The tissue is deformed by simply applying an external force. The second type is dynamic elastography, where waves created by either the imaging system or natural pulsations, caused by for example heartbeats, are tracked. In both types, the response of the tissue to external or internal forces is used to determine its mechanical properties. This is done by obtaining the displacement image, which shows the motion of every sample in the radio frequency (RF) frame during the deformation. We focus on quasi-static freehand strain imaging in this paper, where the strain image is computed by spatially differentiating the deformation field. 

In order to be able to estimate the strain image, we need two RF frames collected before and after applying the external force. One of the problems that free-hand ultrasound elastography faces is the difficulty in choosing suitable RF frames to estimate the strain. If the two RF frames are collected from the same plane and the force is purely axial, they will yield a high-quality strain image. Therefore, the operator needs to be an expert in performing the freehand palpation, rendering this technique very user-dependent. To solve this problem and make the data collection procedure independent of the user's experience, Ranger et al.~\cite{3d_brian_anthony} used a 3D camera to track and compensate any undesired motion that could happen during the data collection. Another approach by both Foroughi et al. \cite{foroughi2013freehand} and Rivaz et al. \cite{rivaz2009tracked} depends on external trackers to collect information about the exact location of the RF frame. By doing this, they can find the RF frames that lie in the same plane, so that they can choose a suitable pair according to some cost function. Aalamifar et al. \cite{robot} used a robot for collecting RF frames. They try to estimate a transformation matrix that transforms the RF frames collected from the robot's tooltip to the ultrasound image frame, using an active echo element.

Although the previously mentioned methods did improve the quality of the strain image, they all need an external device, which complicates the process of data collection and makes it more expensive. Herein, we introduce a novel method using a convolutional neural network (CNN) to determine whether a specific pair of RF frames is suitable for elastography. Although we focus on quasi-static elastography, the method can also be applied to other types of elastography.
\section{METHODS}

In this section, we will discuss data collection for training and testing, and the CNN architecture used. Our model is simply a binary classifier, which is used to determine the suitability of a pair of RF frames for strain estimation. 

Our proposed technique can also be used for automatically finding the best RF frames for a specific pre-selected RF frame. The model achieves that by searching in a window composed of several RF frames (in this work, 8 before and after the pre-specified RF frame).

\subsection{Data Collection}
\label{data_description}
The data used for training and testing the algorithm includes both phantom and \textit{in vivo} data. For the phantom data used in this paper, 4,116 pairs of RF frames were collected at Concordia University's PERFORM Centre from 3 different CIRS phantoms (Norfolk, VA), namely Models 040GSE, 039 and 059 at different locations. 3,290 pairs out of the total data were used for training and validation with a ratio of 80:20, and the remaining data was used for testing.  The ultrasound device used was the 12R Alpinion ultrasound machine (Bothell, WA) with an L3-12H high density linear array probe at a center frequency of 8.5 MHz and sampling frequency of 40 MHz. For the \textit{in vivo} data, 688 pairs of RF frames were collected at Johns Hopkins Hospital from different patients who were undergoing liver ablation for primary or secondary liver cancers. Detailed information about this data is available in \cite{rivaz2011real}. 528 pairs out of the 688 pairs were used for training and validation with a ratio of 80:20, leaving the rest of the pairs for testing. The labelling of the data was done as described in Algorithm 1.

\begin{algorithm}[]
\caption{Labelling the dataset for the CNN classifier}
\begin{algorithmic}[1]
\Procedure {~}{}
\State \text{RF frames $I_{1}$ and $I_{2}$ are passed to PCA-GLUE}
\text{ \hskip 1em \cite{zayed2019fast,hashemi2017global} to obtain the displacement image.}
\State \text{$I_{2}$ is deformed and interpolated according to the}
\text{ \hskip 1em computed displacement image yielding $I_{2}$\textprime.}
\State \text{We partition $I_{1}$ and $I_{2}$\textprime \hskip 0.25em into 9 windows.}
\State \text{Normalized Cross Correlation (NCC) is calculated }
\text{\hskip 1.5em   between every window in $I_{1}$ and its corresponding}
\text{\hskip 1.5em   window in $I_{2}$\textprime, resulting in 9 different NCCs.}
\State \text{The final decision is 1 if \textit{both} the smallest NCC is}
\text{\hskip 1.5em  higher than 0.9 \textit{and} the absolute value of the}
\text{\hskip 1.5em  average displacement is more than 0.5 pixels, and 0}
\text{\hskip 1.5em  otherwise.}
\EndProcedure
\end{algorithmic}
\label{annotating}
\vspace{.12cm}
\vspace{3mm}
\end{algorithm}

It is important to note that steps 2 and 3 in Algorithm 1 are very computationally complex. As such, they cannot be performed in real-time for selecting optimal pairs of RF data. Our proposed method only performs these steps during training, and encodes the results into a computationally efficient CNN.

\subsection{Architecture}
 Suppose we have two RF frames $I_1$ and $I_2$, and we would like to determine the suitability of this pair for strain estimation. We simply input the two frames to the CNN classifier on two different channels, and the output is a binary number 1 or 0. The architecture used is relatively simple as shown Fig.~\ref{CNN_architecture}. Every convolutional layer has a Rectified Linear Unit (ReLU) as the activation function, and is followed by batch normalization. The activation function in the output layer is a softmax, where the output values in the two nodes represent the probability of having a good and a bad pair respectively. The applied optimization technique is the Adam optimizer \cite{kingma2014adam} with a learning rate of $10^{-3}$ and a cross entropy loss function. The CNN code is written in Python using Keras.

\subsection{Training and testing time}
The labelling of the data, which includes applying Algorithm 1 on every single pair of RF frames took 22 hours. Most of this time was spent on displacement estimation (step 2) and interpolating RF data (step 3). 
The actual training of the CNN took 7.4 minutes on a 7th generation 3.4 GHz Intel core i5 desktop with a NVIDIA TITAN V GPU. Inference is very fast, and only takes 5.4 $ms$ to classify two frames of size 2304 by 384. The frames are downsampled by a factor of 2 in the axial direction, to generate smaller input images for the CNN. Note that in comparison, doing steps 2, 3, 5 and 6 in Algorithm 1 for two frames of the same size takes 6.21 seconds, 14.04 seconds, 46.87 $ms$ and 2.45 $ms$ respectively, for a total run-time of 20.3 seconds. In other words, frame selection with CNN is more than 3,700 times faster. It is important to note than CNN computations are performed on a GPU, whereas the steps in Algorithm 1 use a CPU.

\section{EXPERIMENTS AND RESULTS}
In this section, we compare our CNN frame selection method to other methods that choose to pair an RF frame with another by simply skipping one or two frames. We use two metrics for assesing the quality of our classifier, which are accuracy and F1-measure. The accuracy is defined simply as the ratio of the correctly classified instances to the total number of instances given to the classifier. The F1-measure is defined as the harmonic mean of precision and recall.

Fig.~\ref{phantom_results} shows the output of different frame selection methods when tested on one of the phantom datasets. It is clear that our automatic frame selection substantially outperforms the fixed skip frame pairing methods as it chooses more suitable frames, yielding better quality strain images. Table~\ref{table_results} shows the accuracy as well as the F1-measure obtained from our CNN classifier on new phantom datasets, that were not used during training. The results prove the ability of the classifier to generalize to unseen data.
\begin{table}[]
\centering
\vskip 1.5em
\caption{The accuracy and F1-measure of our CNN classifier on the phantom and \textit{in vivo} test data.}\label{tab11}
\begin{center}
\begin{tabular}{lrrrrr}
\toprule
\textbf{Dataset} & \hspace*{-2.5 pt} \textbf{Size} & \textbf{Accuracy}  & \textbf{F1-measure}\\
\midrule
Phantom dataset 1& 228 instances & \hskip 1.5em 96.77\% &\hskip 1.5em 93.68\%\\
Phantom dataset 2& 297 instances &\hskip 1.5em 91.7\% & \hskip 1.5em 89.17\% \\
Phantom dataset 3& 301  instances &\hskip 1.5em 96\% & \hskip 1.5em 96\% \\
\textit{In vivo} dataset& 160  instances &\hskip 1.5em 95.24\% & \hskip 1.5em 92\% \\
\bottomrule
\end{tabular}
\end{center}
\label{table_results}
\end{table}

Fig.~\ref{invivo_results} shows a comparison between the performance of our method and the fixed skip frame pairing on the \textit{in vivo} dataset. Table~\ref{table_results} shows the accuracy as well as the F1-measure obtained from our method. Again, it is clear that our CNN-based method performs substantially better.

\section{CONCLUSION}

In this paper we introduced a new method based on a CNN to automatically choose RF frames that are suitable for strain estimation. Our method is fast, practical and does not need any external hardware. Therefore, it could be used commercially to generate high quality strain images even when used by an inexperienced operator. This can be achieved by simply giving a warning message to the sonographer if the frames used are not suitable.


\addtolength{\textheight}{-12cm} 




\section*{ACKNOWLEDGMENT}

The \textit{in vivo} data was collected at Johns Hopkins Hospital. The authors would like to thank the principal investigators Drs. E. Boctor, M. Choti and G. Hager who provided us with the data. We would like to thank Morteza Mirzaei for providing us with some of the phantom data used in this paper. The authors also acknowledge NVIDIA for donating the graphics card.


\bibliographystyle{IEEEtran}
\bibliography{strings,refs}
\end{document}